\documentclass[pra,aps,twocolumn,amsmath,amssymb]{revtex4}

\usepackage[T1]{fontenc}
\usepackage[english]{babel}    	
\usepackage[pdftex]{graphicx} 	
\usepackage{hyperref}        		
\usepackage{amsmath}
\usepackage{eurosym}
\usepackage{xcolor}
\usepackage{comment}
\usepackage{dsfont}
\usepackage{siunitx}
\begin{document}

\title{Observation and control of quantized scattering halos}

\author{G. Chatelain$^1$, N. Dupont$^1$, M. Arnal$^1$, V. Brunaud$^1$, J. Billy$^1$, B. Peaudecerf$^1$, P. Schlagheck$^2$ and D. Gu\'ery-Odelin$^1$}
\affiliation{
$^1$ Laboratoire Collisions, Agr\'egats, R\'eactivit\'e, IRSAMC, Universit\'e de Toulouse, CNRS, UPS, F-31062 Toulouse, France 
\\
$^2$ CESAM research unit, University of Liege, 4000 Liege, Belgium
}

\begin{abstract}
We investigate the production of $s-$wave scattering halos from collisions between the momentum components of a Bose-Einstein condensate released from an optical lattice. The lattice periodicity translates in a momentum comb responsible for the quantization of the halos' radii. We report on the engineering of those halos through the precise control of the atom dynamics in the lattice: we are able to specifically enhance collision processes with given center-of-mass and relative momenta. In particular, we observe quantized collision halos between opposite momenta components of increasing magnitude, up to 6 times the characteristic momentum scale of the lattice.
\end{abstract}

\maketitle

\emph{Introduction} - Scattering experiments act as a probe revealing at a macroscopic scale the properties of collisional processes that occur at a microscopic scale.  Since the seminal works of H. Geiger, E. Marsden and E. Rutherford \cite{geiger1908,geigermarsden1909,rutherford1911}, such experiments have remained a method of choice to probe atoms, molecules and their interactions.
 In short, the description of scattering in quantum mechanics gives rise to two remarkable features\,: firstly the quantum description of the collision process leads to its decomposition in terms of partial scattering waves \cite{RevModPhys.71.1}, secondly, each of the collisional partners can itself be described as a matter-wave, which can be in a superposition of several components with well defined momenta, leading to multiple elementary collisional processes happening all ``at once". As a result, quantum scattering exemplifies a key feature of quantum mechanics: wave-particle duality.

With the advent of cold atom samples, this topic has been revisited with only a few partial waves involved. The characterization of the collisional properties, and in particular of the $s$-wave scattering length, was performed either using photoassociation measurements \cite{PhysRevLett.74.1315,PhysRevLett.85.1408,Kim_2005,RevModPhys.78.483} or studying the kinetics towards equilibrium of an atomic sample \cite{PhysRevLett.70.414,PhysRevLett.79.625}. In this latter type of measurements, the interplay between partial waves turns out to be subtle: the thermalization rate involves partial waves interferences while the collision rate does not \cite{PhysRevA.73.032706}. Using a 1D collider geometry, the experiments of Refs.~\cite{Buggle2004,Thomas2004} have captured quantum scattering in its purest form: at low energy, the $s$-wave collisions create a spherical shell of pair-correlated atoms, and at slightly higher energy the volume occupied by the scattered atoms reflect the interference between partial waves.  Such collider-like experiments have also been carried out with different species \cite{Thomas_2017, Thomas2018NatCom}. More recently, second- and third-order correlations between momentum-correlated atoms in a collisional halo have been investigated \cite{PhysRevLett.99.150405,PhysRevLett.118.240402}, opening  quantum-nonlocality tests to ensembles of massive particles \cite{Shin2019}.

In this article, we engineer the collisions between atomic ensembles in
a multiple 1D collider, using an out-of-equilibrium Bose-Einstein condensate (BEC) of $^{87}$Rb released from an optical lattice. The collisions occur through $s$-wave scattering, between the discrete momentum components of the BEC, in the course of the time-of-flight, leading to the appearance of scattering halos \cite{Greiner2001}. Through accurate control of the lattice phase and amplitude, we show that it is possible to engineer the dynamics of the BEC in the lattice before the release, and tailor the wavefunction of the atoms after release and expansion, in order to selectively enhance the collision processes between specific momentum components of the BEC. We can thus produce quantized scattering halos from collisions with a chosen relative and/or center-of-mass (c-o-m) momentum. 

\begin{figure}
	\includegraphics[width=\linewidth]{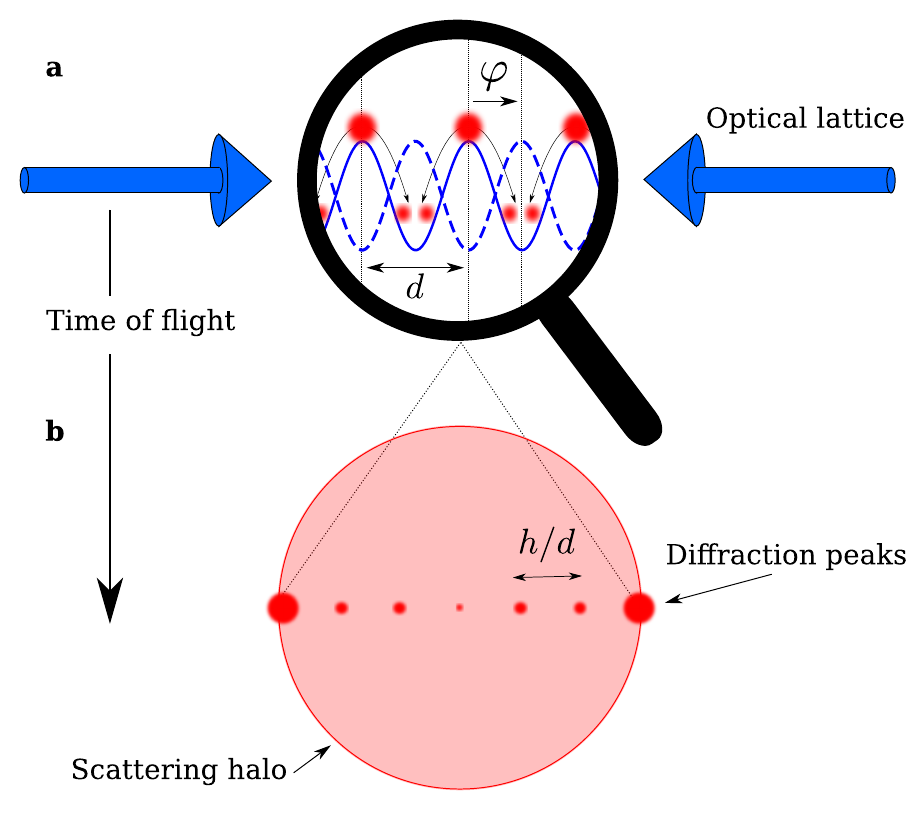}
	\caption{Scheme of the experimental protocol.  A Bose-Einstein condensate (red clouds) is initially loaded in an optical lattice (dashed blue line). {\bf a} The lattice is suddenly phase-shifted by an amount $\varphi=\varphi_0$ (solid blue line); here $\varphi_0=\ang{180}$. The subsequent dynamics in the lattice modifies the momentum distribution. {\bf b} After a sufficiently long time-of-flight, the momentum distribution exhibits diffraction peaks, separated by $\hbar k_L=h/d$, characteristic of the wave nature of the BEC in the lattice (red disks) along with isotropic scattering halos due to $s$-wave scattering between atoms (light red disk).}
	\label{fig:principe}
\end{figure}

\begin{figure*}[ht!]
	\includegraphics{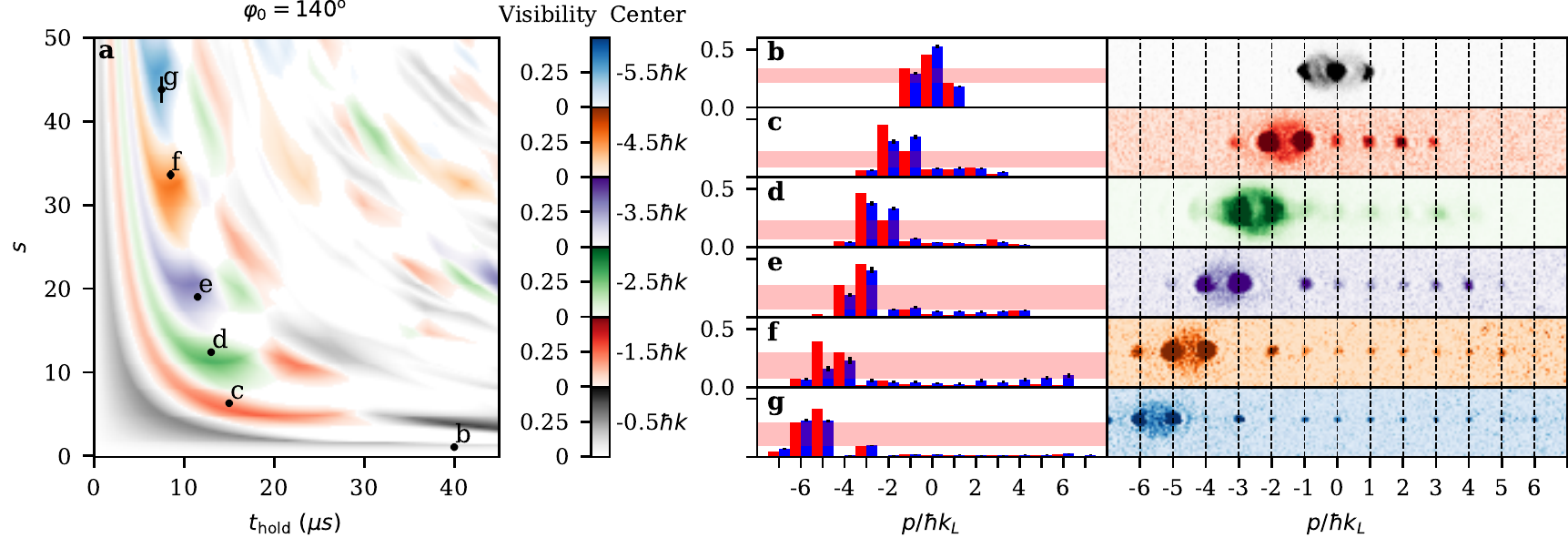}
	\caption{{\bf Collision halos of $1\times h/d$ diameter and varying center-of-mass momentum.} {\bf a.} Simulated visibility $V$ of the orders of diffraction (see text) as a function of the lattice depth $s$, and holding time $t_{\textrm{hold}}$, for a sudden phase shift of $\varphi_0 = \ang{140}$. The center-of-mass momentum of the expected dominant collision halo is color-coded. Black markers indicate experimentally tested parameters, with the vertical error bar showing the standard deviation of the independent measurement of $s$. {\bf b-g} Left: experimental diffraction orders probability distributions $\{\pi_j\}$ (blue, offset to the right), with black error bars showing one standard deviation, and simulated ones for the same parameters $\varphi_0,\, s,\, t_{\textrm{hold}}$ (red, offset to the left). The red shaded horizontal bands represent the visibilities for the theoretically simulated distributions, and stretch between the probabilities of the second and third most populated diffraction orders. Right: single-shot experimental absorption images from which the momentum distributions are extracted. The collision halo between the two most populated orders are clearly visible. The color code of the absorption images indicates the c-o-m momentum for the scattering halo, with the same color code as Fig.~\ref{fig:com}{\bf a}. The color scale is set to enhance the collision halos (clipping high values on some diffraction orders). The parameters used for data {\bf b-g} are $\{s,t_{\textrm{hold} } [\si{\micro\second}]\}=\{1.01\pm0.02,40\},\{6.35\pm0.23,15\},\{12.67\pm0.26,14\},\{18.67\pm0.26,11.5\},\{33.15\pm0.51,8.5\}$ and $\{43.50\pm1.61,7.5\}$, respectively.}
	\label{fig:com}
\end{figure*}

\emph{Background} - When a BEC is loaded in an optical lattice, the periodic structure of the lattice imprints on the wavefunction. In particular the resulting momentum distribution is made up of equally spaced peaks, separated by an interval $h/d$, where $d$ is the lattice spacing and $h$ the Planck constant. The shape of each individual peak is set by atomic interactions and
the extra confinement superimposed to the lattice.
This momentum distribution can be measured by releasing the atoms from the lattice and allowing the atomic cloud to ballistically expand for a sufficiently long time-of-flight (TOF), $t_{TOF}$: the spatial density $n(\mathbf{r})$ then reproduces the initial momentum density $\tilde{n}(\mathbf{p})$ up to a scaling: $n(\mathbf{r},t_{TOF})=\tilde{n}(\mathbf{p}=m\mathbf{r}/t_{TOF},t=0)$, with $m$ the mass of the atom (see Fig.~\ref{fig:principe}).

However, this comb pattern resulting from the wave nature of the BEC is only a partial description of the final atomic density distribution: for interacting atoms, collisions originating from the particle nature of matter may occur during the ballistic expansion. In the low-energy regime characteristic of ultracold atoms, $s$-wave elastic scattering dominates which, due to energy and momentum conservation in the collision process, results in spherical halos centered on the center of mass of the two colliding atomic wavepackets (see \cite{Band2000,Zin2005} and  Appendix \ref{appendixA}). 

Thus, spherical collision halos are expected to appear in the momentum distribution between the diffraction orders after time-of-flight, as depicted in Figure~\ref{fig:principe}. Each diffraction order, centered on a position $x_j=j\times h/(md) \times t_{TOF}\ (j\in\mathds{Z})$, contains a fraction $\pi_j$ of the number of atoms $N$ initially in the BEC. These fractions determine the atomic density in the halos : in a perturbative approach, the number of collisions between orders $j$ and $k$, and therefore the number of atoms scattered in the corresponding halo, is proportional to the product $\pi_j \pi_k$ (\cite{Tenart2020}, Appendix \ref{appendixA}).

In this work, we \emph{engineer} the (dominant) scattering halos, by precisely controlling the state of the BEC in the lattice before release. To this end, we apply a sudden displacement to the lattice on a scale smaller than the lattice spacing. This triggers out-of-equilibrium dynamics of the BEC inside the lattice. Controlling the duration of this evolution in the lattice prior to release (see Fig.~\ref{fig:principe}), we effectively tailor the momentum distribution $\{\pi_j\}$. 

\medskip

\begin{figure*}[ht!]
	\includegraphics{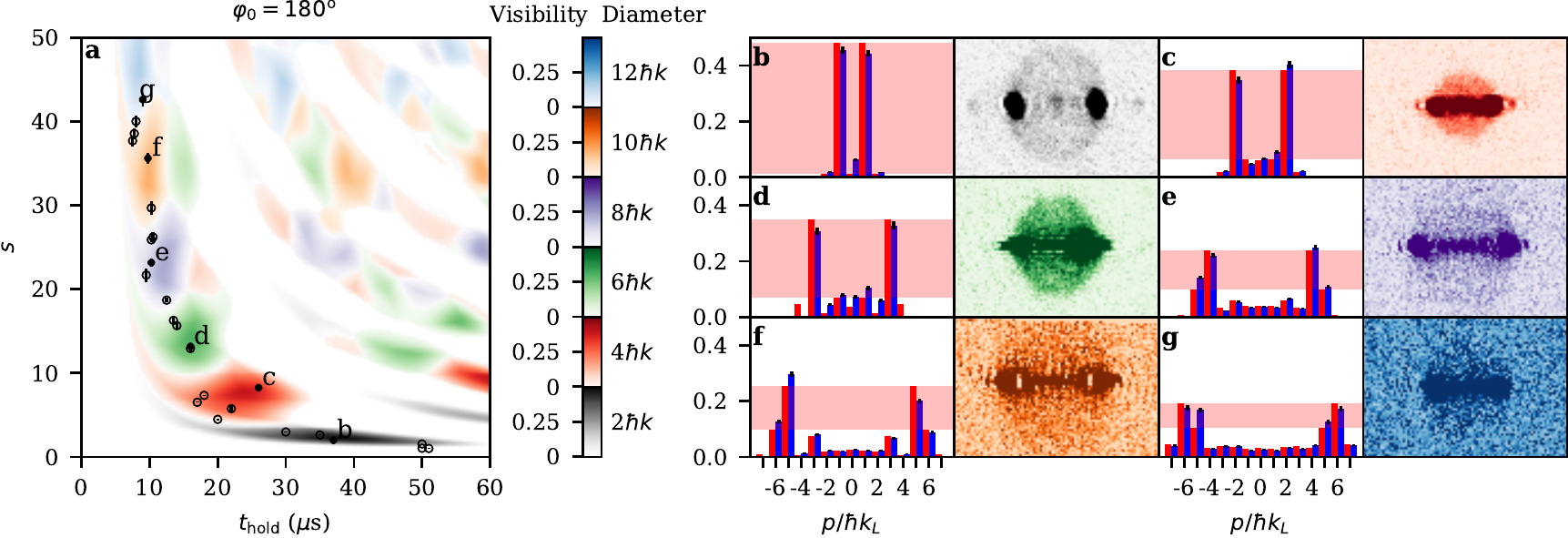}
	\caption{{\bf Collision halos of $2 j\times h/d$ diameter $(1\leq j \leq 6)$ and center-of-mass momentum 0.} {\bf a.} Simulated visibility $V$ of the orders of diffraction (see text) as a function of the lattice depth $s$, and holding time $t_{\textrm{hold}}$, for a shift of $\varphi_0 = \ang{180}$. The c-o-m momentum of the collision is zero, and the relative momentum is color-coded. Black markers indicate experimentally tested parameters, with the vertical error bar showing the standard deviation of the independent measurement of $s$. Full symbols relate to the data shown in {\bf b-g}, and all symbols including empty ones relate to figure \ref{fig:quantification}{\bf d}. {\bf b-g} Left: experimental diffraction orders probability distributions $\{\pi_j\}$ (blue, offset to the right), with black error bars showing one standard deviation, and simulated ones for the same parameters $\varphi_0,\, s,\, t_{\textrm{hold}}$ (red, offset to the left). The red shaded horizontal bands represent the visibilities for the theoretically simulated distributions, and stretch between the probabilities of the second and third most populated diffraction orders. Right: corresponding single-shot experimental absorption images. The collision halos between the two most populated orders are clearly visible. The color code of the absorption images indicates the diameter for the scattering halo, with the same code as Fig.~\ref{fig:diameter}{\bf a}. The color scale is set to enhance the collision halos (clipping high values on some diffraction orders). The parameters used for data {\bf b-g} are $\{s,t_{\text{hold}}  [\si{\micro\second}]\}=\{1.99\pm0.03,37\},\{8.26\pm0.09,26\},\{13.18\pm0.33,16\},\{23.17\pm0.45,10.2\},\{35.61\pm0.61,9.8\}$ and $\{40.02\pm0.69,8.0\}$, respectively.}
	\label{fig:diameter}
\end{figure*}

\emph{Methods} - We perform our experiments in a hybrid trap \cite{fortun2016} in which we obtain pure rubidium-87 Bose-Einstein condensates of $2\cdot 10^5$ atoms in the lowest hyperfine state $|F=1, m_F = -1\rangle$. These BECs are loaded in a one-dimensional optical lattice produced by two counterpropagating laser beams of wavelength $\lambda=1064$ nm superimposed to the optical dipole beam of the hybrid trap. In the optical lattice, the atoms experience the following potential :

\begin{align}
	U(x,t) = -\frac{s}{2}E_L\cos\left(k_L x + \varphi\right), 
	\label{eq:potential}
\end{align}
where $E_L = \hbar^2k_L^2/(2m)$ ($E_L=4E_R$, with $E_R$ the recoil energy) and $k_L = 2\pi/d$ are respectively the energy and the wavevector associated to the lattice. 
The dimensionless depth of the lattice $s$ is independently and precisely calibrated \cite{CabreraCalibration} for each experiment presented here. 
The phase $\varphi$ is set by the relative phase between the two phase-locked acousto-optic modulators controlling the lattice beams. In $^{87}$Rb, higher-order $d-$wave collisions occur for energies of $\sim \SI{200}{\micro\kelvin}$ \cite{Buggle2004,Thomas2004}, while in the experiments shown here we impart at most $\sim \SI{25}{\micro\kelvin}$ of collisional energy. We are therefore in the purely $s$-wave scattering regime. 
Within this regime, in order for the changes in momentum distribution to directly relate to the number of atoms in the halos, the scattering cross section needs to be independent from the relative speed across all our experiments (see Appendix \ref{appendixA}), and not be affected by scattering suppression due to the superfluid onset \cite{chikkatur2000}. We have checked that the minimum relative velocity we can impart in a collision is of $h/(md)=\SI{8.6}{\milli\meter\per\second}$, which is much larger than the peak sound velocity for the BEC in one of the colliding diffraction orders ($c\sim\SI{1.5}{\milli\meter\per\second}$). Therefore the scattering cross-section is the same for all collision halos presented here.

In order to tailor the momentum distribution after expansion, we use the following general procedure: we first load adiabatically the BEC for $\varphi=0$; then a sudden phase shift $\varphi_0$ is applied \footnote{A shift $\varphi_0$ leads to a displacement of the lattice by $\delta x=d\times\frac{\varphi_0}{2\pi}$}, triggering out-of-equilibrium dynamics ($t=0^+$); in the following holding time, the momentum distribution $\{\pi_j(t)\}$ evolves in the lattice; finally, the lattice is released at $t=t_{\textrm{hold}}$. The atom cloud then expands for a duration $t_{TOF}$, and an absorption image of the resulting density $n(\mathbf{r})$ is taken to measure the  momentum distribution and the scattering halos.

As a guide to our engineering, we numerically compute the evolution of a one-body wavefunction, initially in the ground state of an infinite optical lattice of depth $s$. After the phase shift $\varphi_0$, the wavefunction is decomposed on the lattice bands, and evolves during the holding time $t_{\textrm{hold}}$. We then extract the final momentum distribution (see for example Fig~\ref{fig:com}{\bf b-g}, red bars and Appendix \ref{appendixC}).

\medskip

\emph{Results} - In a first series of experiments, we demonstrate control of the c-o-m momentum of the scattering halo. For that purpose, we use the phase shift $\varphi_0=\ang{140}$, which puts the atom clouds in the lattice on the side slope of each well (see Fig.~\ref{fig:principe}). As the atoms reach the bottom of the well in the following dynamics, we expect a high c-o-m momentum, which is higher the deeper the lattice is, with a small dispersion of the distribution over diffraction orders.  

As the dominant collision halo will occur between the two mostly populated diffraction orders, we define a \emph{visibility} parameter $V$ as \emph{the difference between the fractions $\pi_j$ of the second and third most populated diffraction orders in the momentum distribution}. It is a measure of how well the two diffraction orders contributing to the main halo "stand out", and varies between 0 (the third highest order is as populated as the second) and 0.5 (the two orders contributing to the main halo are the only ones populated). In Fig.~\ref{fig:com}{\bf a}, the values of the visibility $V$ obtained from the wavefunction simulation are plotted over a large range of lattice depths $s$ and holding times $t_{\textrm{hold}}$. As we are here interested in controlling mainly the c-o-m momentum, the visibility $V$ is only plotted if the two dominant orders are next to each other (separated by $1\times h/d$ in momentum), and is otherwise set to 0. Finally we also indicate the value of the c-o-m momentum by a color code. As intuitively expected, higher c-o-m momenta can be reached for deeper lattices, for a holding time that gets shorter the deeper the lattice gets.

At a given depth, our analysis in terms of visibility shows that "patches" of lower c-o-m momentum values can be obtained for longer holding times. However the visibility in these patches is less pronounced, 
meaning there are likely more than two significantly populated orders of diffraction for these parameters, and we can expect that several collisional processes will occur simultaneously in the experiment. Therefore
we have not represented experimental data for these values.

In Fig.~\ref{fig:com}{\bf b-g}, we represent a few snapshots of the density distributions obtained after time-of-flight for parameters that are indicated in Fig.~\ref{fig:com}{\bf a}. For each of these measurements, the lattice depth was calibrated independently \cite{CabreraCalibration}. The disk-shaped halo due to collisions between the two most populated orders, separated by $1\times h/d$, is clearly visible, and is the main feature besides the regular structure of the diffraction orders. Alongside the absorption pictures we represent the measured histograms $\{\pi_j\}$ of the populations in the diffraction order $p_j=j\times h/d$. These are compared to the calculated histograms from the wavefunction simulation, and the visibility parameter from the simulation is indicated by a red-shaded area. Note that the sign of the c-o-m momentum can be easily changed by changing the sign of the shift $\varphi_0$.

\begin{figure*}[ht!]
	\includegraphics{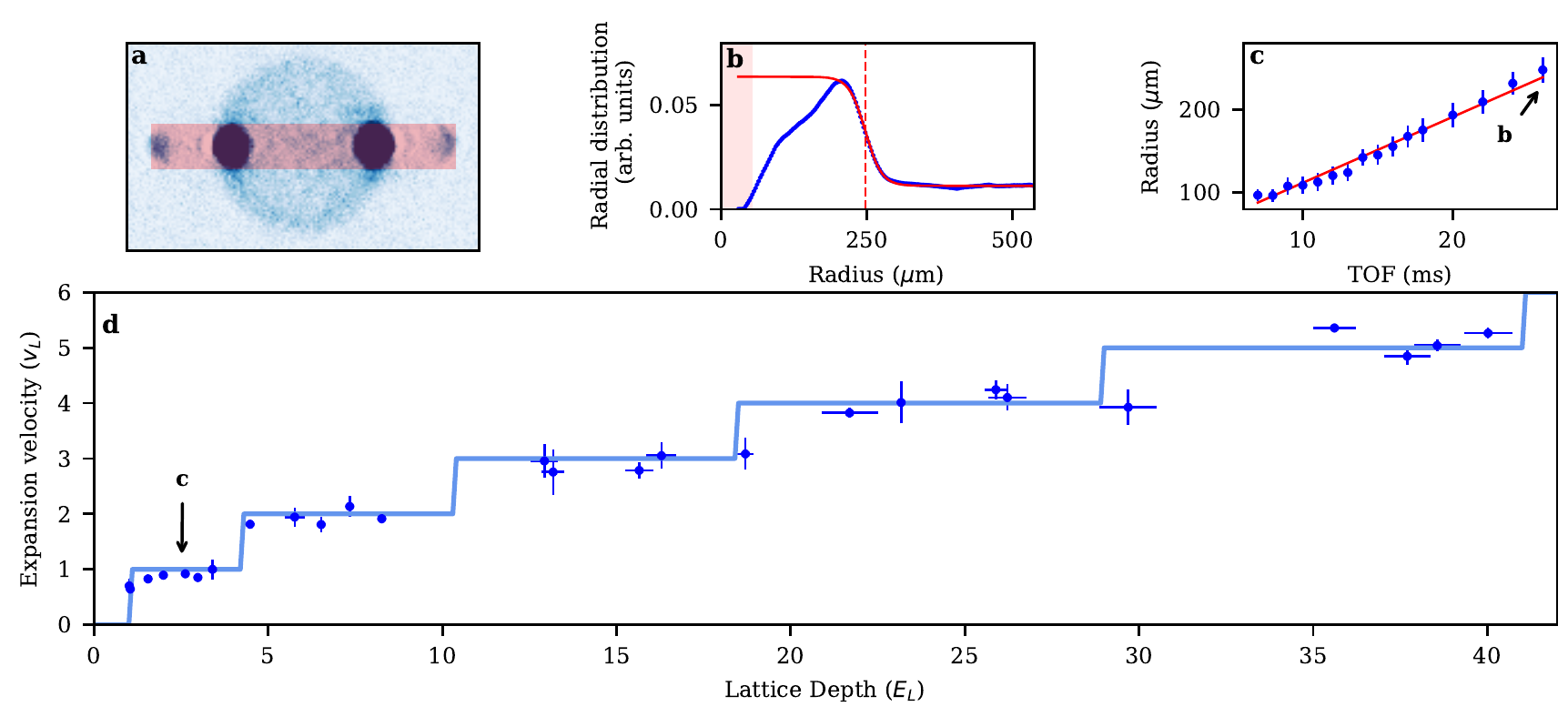}
	\caption{{\bf Quantization of the scattering halo radius.} {\bf a} Typical absorption image from which the radius of the main collision halo is measured. The red rectangle indicates the mask used to hide the diffraction orders in order to extract the halo's characteristics. {\bf b} From the masked image {\bf a}, an angular average is performed (blue markers) and a sigmoid fit (red line) to the resulting radial distribution allows us to extract the radius of the collisional halo (dashed red line). {\bf c} The procedure shown in {\bf b} is repeated for multiple TOF values (blue markers). A linear fit (solid red line) yields the velocity expansion of the scattering halo. {\bf d} Measured expansion velocities as a function of the lattice depth $s$ (blue markers). The parameters $\{s,t_{\textrm{hold}}\}$ corresponding to these data points are indicated by the disks in Fig.~\ref{fig:diameter}{\bf a}. The quantization of the expansion velocity of the main scattering halo  in terms of $v_L=h/(md)$ is apparent. The blue line indicates the expected expansion velocity of the halo originating from the diffraction orders with the highest visibility at depth $s$ (see text). The model and the observed steps are in very good agreement.}
	\label{fig:quantification}
\end{figure*}

Even though the fraction of atoms in the halo can get rather large (see below), the agreement between the simulated distributions, which do not include interactions, and the measured distributions, for which collisions have occurred, is very good. This can seem surprising, as atoms are removed from the diffraction peaks during the collisions. However in any collision between two orders of diffraction, an equal number of atoms is removed from the involved peaks, after which the measured distribution (as shown in blue in Fig.~\ref{fig:com}) is the new normalized distribution in the peaks. We have simulated that even with a fraction of colliding atoms as high as 30\%, the expected change between the non-interacting theoretical distribution and the distribution obtained after accounting for the removal of the colliding atoms and normalizing remains small (<5\%). This means that the one-body wavefunction simulation is a surprisingly robust guide in our investigation of collision halos, as is further evidenced below (see Fig.~\ref{fig:quantification}).

In a second series of experiments, our aim was to control specifically the relative momentum of the collisions. To that end, we apply a shift $\varphi_0=\ang{180}$: the atom clouds in the lattice are placed at the top of the lattice potential, and split in two identical clouds that fall on either side (see Fig.~\ref{fig:principe}). In a classical picture, after a holding time that allows the split clouds to reach the bottom of the wells, momentum orders of opposite sign should be equally populated, with zero c-o-m momentum. Their typical magnitude should also increase with lattice depth $s$.

In Fig.~\ref{fig:diameter}{\bf a}, we plot the values of the visibility $V$, as defined before, over a range of values of lattice depths $s$ and holding times $t_{\textrm{hold}}$, for $\varphi_0=\ang{180}$, as calculated from the wavefunction simulation. We do not impose a condition anymore on the separation between the most populated orders (note that in this case, the theoretical distributions are symmetrical and the two highest peaks are equally populated). The diameter of the expected collision halo (with zero c-o-m momentum) is indicated by a color code. As expected, the relative collisional momentum increases with lattice depth. In the main feature of this figure, the relative momentum of the most populated orders increases in steps of $2\times h/d$ as the lattice depth is increased, for decreasing holding times (a behaviour that is expected intuitively).

In Fig.~\ref{fig:diameter}{\bf b-g} we represent snapshots of the density distributions obtained after time-of-flight for parameters that are indicated in Fig.~\ref{fig:diameter}{\bf a}. Disk-shaped collision halos of increasing diameters, quantized in units of $2h/d$, are clearly visible. To the left of the absorption images, we represent the momentum distributions ${\pi_j}$ as extracted from absorption images. We compare the experimentally measured histograms with those from the wavefunction simulation for the same parameters $s$ and $t_{\textrm{hold}}$, and find once again very good agreement. Since large collision halos become very dilute at the long time-of-flight needed for the reconstruction of the histograms (a few tens of \si{\milli\second}), we used a different time-of-flight for the visualization of the halos (typically a few \si{\milli\second}). 

The observed fraction of colliding atoms we measure is in general larger than predicted by a perturbative theory (see Appendix \ref{appendixA}). 
To measure it, we count the atoms in the collision halos when the diffraction orders are masked, as in Fig.~\ref{fig:quantification}{\bf a}. We account for the portion of disk hidden, and calculate the ratio of the number of atoms in the halo to the sum of the numbers in the halo and in the diffraction orders. 
We measure typically a fraction of $30\pm3\%$ in the data presented Fig.~\ref{fig:com}{\bf b} and Fig.~\ref{fig:diameter}{\bf b} (from a statistical average of the fractions in 15 absorption images), where the theory predicts 20\%. This is in contrast to the results in \cite{Tenart2020} which were well described with a similar prediction; but while these results involved typically a few percent of the atoms of the BEC in the collisions between two orders, we observe collisions between highly populated orders, and get typically five times larger atom fractions involved in the collisions, at which point a non-perturbative approach may be needed to get quantitative agreement. We also independently checked that
this higher number is not due to collisions happening inside the lattice during the non-equilibrium evolution time $t_{\textrm{hold}}$ (see Appendix \ref{appendixB}).

We have also verified that the diameter of the dominant collision halo in momentum space is indeed quantized in terms of $h/d$ (the momentum scale set by the lattice), and that larger diameters are reached for larger lattice depths. This analysis was performed independently from the known position of the diffraction orders, or any a priori visibility calculation. This is also performed for a phase shift of $\varphi_0=\ang{180}$, and for the experimental parameter values shown in Fig.~\ref{fig:diameter}{\bf a}. For each of the parameter values $\{s$,$t_{\textrm{hold}}\}$, we record a sequence of images with increasing  TOF. On each of these images, as in Fig.~\ref{fig:quantification}{\bf a}, we mask the diffraction orders, to focus on the most visible disk-shaped halo. By performing an angular average, we obtain a radial distribution, clearly showing the edge of the collision halo. We fit the position of this edge with a sigmoid function (Fig.~ \ref{fig:quantification}{\bf b}), which gives us the radius of the scattering halo for a given TOF. We then extract the speed of expansion of the scattering halo by fitting the linear growth of the radius with TOF (Fig.~\ref{fig:quantification}{\bf c}).

This procedure was repeated for multiple values of $s$, choosing for each depth a holding time $t_{\textrm{hold}}$ yielding a scattering halo with the strongest signal. The results are shown in Fig.~\ref{fig:quantification}{\bf d}. We find indeed that the expansion velocity of the dominant halo, as measured from the halo only, shows sharp jumps between discrete values that are an integer multiple of $v_L=h/(md)$, the velocity scale set by the lattice. As the lattice depth increases, so does the collisional energy, which is converted from the lattice potential energy during $t_{\textrm{hold}}$, and so does the expansion velocity of the dominating halo. Alongside the experimental data in Fig.~\ref{fig:quantification}{\bf d}, we represent the expected collision velocity for the dominant scattering halo, as given by the point with highest visibility for a given depth $s$ on the map Fig.~\ref{fig:diameter}{\bf a}, obtained from the wavefunction simulation. We find that the experimentally observed steps for the relative collision velocity in the dominant halos are in remarkably good agreement with the characterization in terms of visibility of the diffraction orders.

\begin{figure*}[ht!]
	\includegraphics{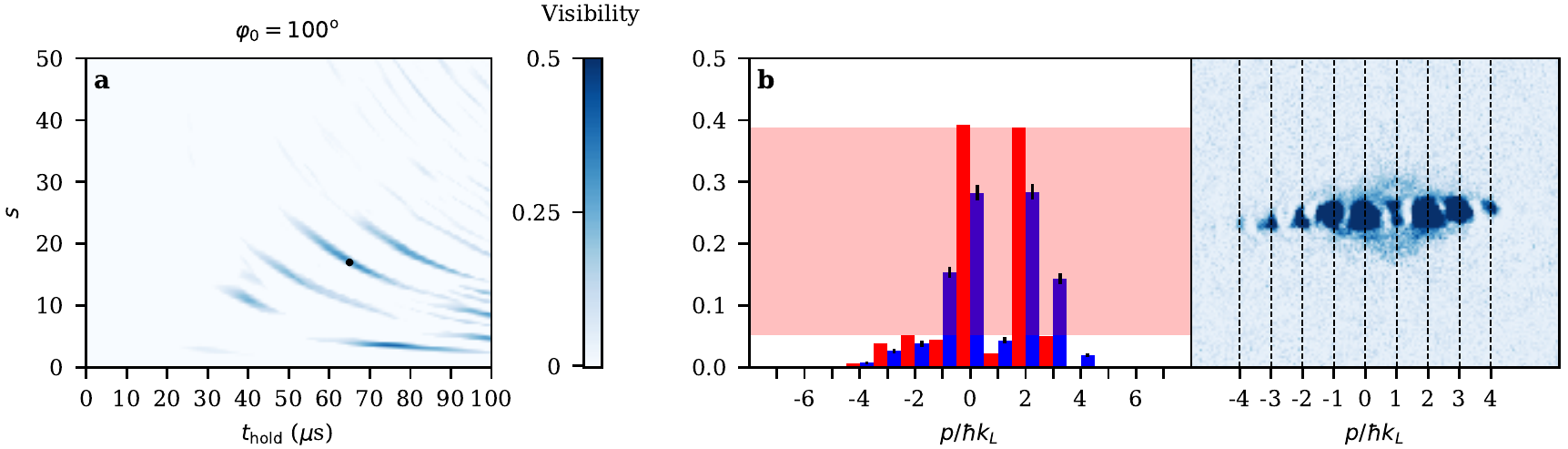}
	\caption{\textbf{ Collision halo of $2\times h/d$ diameter and $1\times h/d$ c-o-m momentum.} {\bf a.} Simulated visibility $V$ of the orders of diffraction (see text), with the added constraint that the two most populated orders have momenta $0\times h/d$ and $2\times h/d$, as a function of the lattice depth $s$, and holding time $t_{\textrm{\tiny{hold}}}$, for a shift of $\varphi_0 = 100^{\textrm{\tiny{o}}}$. The black marker indicates the experimentally tested parameters, with the vertical error bar showing the standard deviation of the independent measurement of $s$. \textbf{b} Left: experimental diffraction orders probability distribution $\{ \pi_j \}$ (blue, offset to the right), with black error bars showing one standard deviation, and simulated one for the same parameters $\varphi_0$, $s$, $t_{\textrm{\tiny{hold}}}$ (red, offset to the left). The red shaded horizontal band represents the visibility for the theoretically simulated distribution, and stretches between the probabilities of the second and third most populated diffraction orders. Right: corresponding single-shot experimental absorption image. The collision halo between the two most populated orders is clearly visible. The color scale is set to enhance the collision halo (clipping high values on some diffraction orders). The parameters used for data \textbf{b} are $\{ s, t_{\textrm{\tiny{hold}}} \left[\mu s\right] \} = \{ 17.93 \pm 0.27, 65.25 \}$.}
	\label{fig:both}
\end{figure*}

\emph{Discussion and conclusion} - 
In this work, we have demonstrated that we can engineer at will the collision halos from ultracold atoms released from an optical lattice. This is achieved through the control of the dynamics of a BEC in the optical lattice, which allows us to tailor the momentum distribution giving rise to the collisions after release from the trap. We have shown that we can selectively populate diffraction orders to impart a large momentum either to the center-of-mass motion of colliding atom clouds, or to their relative motion. The search for appropriate experimental parameters has been guided by the simulation of the dynamics of a non-interacting BEC in the lattice, and the identification of two highly populated diffraction orders with the desired characteristics. The energies involved in the collisions are quantized in terms of the lattice characteristic momentum but can be adjusted through a wide range of values.

The protocol used here to demonstrate separate control on center-of-mass and relative momentum can be extended to realize arbitrary combinations of the two. In Fig.~\ref{fig:both}, we provide an example where we control simultaneously the collision halo diameter ($2\times h/d$) and the  center-of-mass momentum, ($1\times h/d$). This still relies on the simple control of the static lattice depth and a sudden phase shift, and a visbility search for the desired diffraction orders. 
This work could be extended further using more elaborate dedicated schemes for the control of the momentum distribution, by an appropriate time-dependent shaping of both the depth and phase of the lattice before time-of-flight expansion. 
This could provide a fairly simple and general technique alongside more conventional atom-optics schemes \cite{cronin2009}. Such an optimization could also be of interest for matter-wave interferometry where the engineering of momentum superpositions of atomic ensembles is a key technique \cite{bongs2019,dgocct}.

\acknowledgments
This work was supported by Programme Investissements d'Avenir under the program ANR-11-IDEX-0002-02, reference ANR-10-LABX-0037-NEXT, and research funding Grant No. ANR-17-CE30-0024. M.A. acknowledges support from the DGA (Direction G\'en\'erale de l'Armement), and N.D. support from R\'egion Occitanie and Universit\'e Paul Sabatier.

\begin{appendix}

\newcommand{\vek}[1]{\mathbf{#1}}
\newcommand{\veg}[1]{\boldsymbol{#1}}
\newcommand{\bra}[1]{\langle#1|}
\newcommand{\ket}[1]{|#1\rangle}
\newcommand{\bracket}[2]{\langle#1|#2\rangle}

\section{Perturbative theory of scattering halos}
\label{appendixA}

In this section, we provide a quantitative theoretical description of the atom-atom scattering processes giving rise to the halo structures that are observed in the experiments.
The perturbative framework that we develop is inspired from previous works that were investigating colliding Bose-Einstein condensates \cite{Band2000,Zin2006}.
The results that it yields are consistent with the calculations undertaken in Ref.~\cite{Tenart2020}.

Our starting point is a perfect Bose-Einstein condensate that is prepared within a one-dimensional optical lattice.
The atoms of the condensate are sharing the single-particle wavefunction
\begin{equation}
  \Phi_0(\vek{r}) = \phi(\vek{r}) \sum_{l=-\infty}^\infty \psi_l e^{i l k_L z} \label{eq:phi0}
\end{equation}
with $\vek{r} \equiv (x,y,z)$ and the lattice being oriented along the $z$ axis of the coordinate system.
$\psi_l$ are the Fourier components of the periodic condensate wavefunction in the lattice. The function
\begin{equation}
  \phi(\vek{r}) = \frac{1}{\sqrt{R}^3} \varphi(\vek{r}/R) \label{eq:phi}
\end{equation}
with
\begin{equation}
  \varphi(\veg{\rho}) = \sqrt{\frac{15}{8\pi} \left( 1 - \frac{\omega_\perp^2}{\bar{\omega}^2} (\rho_x^2 + \rho_y^2) - \frac{\omega_{||}^2}{\bar{\omega}^2} \rho_z^2\right)} \label{eq:TF}
\end{equation}
is a Thomas-Fermi envelope that accounts for the presence of a weak overall elliptic confinement with the longitudinal and transverse frequencies $\omega_{||}$ and $\omega_\perp$, respectively, and with $\bar{\omega} = (\omega_{||} \omega_\perp^2)^{1/3}$.
The associated Thomas-Fermi radius $R$ is straightforwardly evaluated as
\begin{equation}
  R = (15 N a_s \bar{a}^4)^{1/5}
\end{equation}
with $\bar{a} = [\hbar/(m\bar{\omega})]^{1/2}$, where $N$ is the number of atoms in the condensate and $a_s$ denotes the $s$-wave scattering length.

At time $t=0$ the lattice and confinement potentials are switched off and the atomic cloud is thereby allowed to freely expand.
If atom-atom interactions could be completely neglected from that instant on, the momentum distribution of the atoms would be simply given by the Fourier transform of the condensate wavefunction, i.e., we would have
$n_0(\vek{p},t) = N |\tilde{\Phi}_0(\vek{p})|^2$ with
\begin{equation}
  \tilde{\Phi}_0(\vek{p}) = \sum_{l=-\infty}^{\infty} \psi_l \tilde{\phi}(\vek{p}-l \hbar k_L \vek{e}_z) \label{eq:p0}
\end{equation}
where
\begin{equation}
  \tilde{\phi}(\vek{p}) = \frac{1}{\sqrt{2\pi \hbar}^3} \int d^3r \phi(\vek{r}) e^{- i \vek{p}\cdot\vek{r}/\hbar}
\end{equation}
is the Fourier transform of the Thomas Fermi profile \eqref{eq:phi}.
As we generally have $k_L R \gg 1$ in the experiment, Eq.~\eqref{eq:p0} suggests the appearance of tighly localized momentum density peaks centered about integer multiples of the lattice momentum $l \hbar k_L$, each one of those populated with $N |\psi_l|^2$ atoms.

The key approximation that we undertake here is to assume that we can account for the presence of atom-atom interaction during the free expansion process of the condensate in a perturbative manner, using first order quantum perturbation theory.
Employing the interaction representation, the time evolution of the many-body state $\ket{\Psi_t}$ describing the atomic cloud is approximately written as
\begin{equation}
  \ket{\Psi_t} \simeq \ket{\Psi_0} - \frac{i}{\hbar} \int_0^t dt' \hat{U}(t') \ket{\Psi_0} + O(a_s^2) \,.
\end{equation}
in linear order in the $s$-wave scattering length $a_s$.
Here,
\begin{eqnarray}
  \hat{U}(t) & = & \frac{g}{2(2\pi\hbar)^3} \int d^3 p_1 \int d^3 p_2 \int d^3 p_1' \int d^3 p_2' \nonumber \\
  && \times \delta\left( \vek{p}_1 + \vek{p}_1' - \vek{p}_2 - \vek{p}_2' \right) \nonumber \\
  && \times \hat{\psi}_t^\dagger(\vek{p}_2) \hat{\psi}_t^\dagger(\vek{p}_2') \hat{\psi}_t(\vek{p}_1') \hat{\psi}_t(\vek{p}_1) \label{eq:Uo}
\end{eqnarray}
with $g = 4\pi \hbar^2 a_s / m$ is the two-body interaction Hamiltonian expressed in terms of the atomic creation and annihilation operators in momentum space which evolve according to
$\hat{\psi}_t^\dagger(\vek{p}) = \hat{\psi}^\dagger(\vek{p}) \exp[i t p^2/(2 m \hbar)]$ as well as $\hat{\psi}_t(\vek{p}) = \hat{\psi}(\vek{p}) \exp[-i t p^2/(2 m \hbar)]$, respectively, and which fulfill the bosonic commutation relation
\begin{equation}
  \left[\hat{\psi}_t(\vek{p}),\hat{\psi}_t^\dagger(\vek{p}')\right] = \left[\hat{\psi}(\vek{p}),\hat{\psi}^\dagger(\vek{p}')\right] = \delta(\vek{p} - \vek{p}') \,.
\end{equation}
We can therefore rewrite Eq.~\eqref{eq:Uo} as
\begin{eqnarray}
  \hat{U}(t) & = & \frac{a_s}{4\pi^2m \hbar} \int d^3 p_1 \int d^3 p_2 \int d^3 p_1' \int d^3 p_2' \nonumber \\ && \times \delta\left( \vek{p}_1 + \vek{p}_1' - \vek{p}_2 - \vek{p}_2' \right) \nonumber \\ && \times \exp\left[\frac{i t}{m \hbar}(\vek{p}_1-\vek{p}_2)\cdot(\vek{p}_1'-\vek{p}_2)\right] \nonumber \\ && \times \hat{\psi}^\dagger(\vek{p}_2) \hat{\psi}^\dagger(\vek{p}_2') \hat{\psi}(\vek{p}_1') \hat{\psi}(\vek{p}_1) \label{eq:UI}
\end{eqnarray}
and use the standard definition of the momentum-space field operators giving rise to
\begin{equation}
  \hat{\psi}(\vek{p}) \ket{\Psi_0} = \tilde{\Phi}_0(\vek{p}) \hat{b}_0 \ket{\Psi_0} \label{eq:b0}
\end{equation}
where $\hat{b}_0$ is the annihilation operator associated with the condensate orbital.

In the following, we focus on regions in momentum space located far away from the lattice density peaks $|\tilde{\phi}(\vek{p}-l \hbar k_L \vek{e}_z)|^2$ which we obtained in zeroth order in the interaction strength.
For a momentum $\vek{p}$ being in such a scarcely populated region we can safely set $\tilde{\Phi}_0(\vek{p}) = 0$ and hence $\hat{\psi}(\vek{p}) \ket{\Psi_0} = 0$.
The atom density detected at such a momentum is therefore determined as
\begin{equation}
  n(\vek{p},t) = \bra{\Psi_t} \hat{\psi}^\dagger(\vek{p}) \hat{\psi}(\vek{p}) \ket{\Psi_t} \simeq \bracket{\Pi_t(\vek{p})}{\Pi_t(\vek{p})} \label{eq:np}
\end{equation}
with
\begin{equation}
  \ket{\Pi_t(\vek{p})} = - \frac{i}{\hbar} \int_0^t dt' \left[ \hat{\psi}(\vek{p}), \hat{U}(t') \right] \ket{\Psi_0}
\end{equation}
in lowest nonvanishing order in the $s$-wave scattering length.
Evaluating
\begin{eqnarray}
  \left[ \hat{\psi}(\vek{p}), \hat{U}(t') \right] & = & \frac{a_s}{2\pi^2m \hbar} \int d^3 p_1 \int d^3 p_1' \nonumber \\ && \times \exp\left[\frac{i t'}{m \hbar}(\vek{p}_1-\vek{p})\cdot(\vek{p}_1'-\vek{p})\right] \nonumber \\ && \times \hat{\psi}^\dagger(\vek{p}_1+\vek{p}_1'-\vek{p}) \hat{\psi}(\vek{p}_1') \hat{\psi}(\vek{p}_1)
\end{eqnarray}
and using Eq.~\eqref{eq:b0} in combination with Eq.~\eqref{eq:p0} , we obtain the expression
\begin{eqnarray}
\ket{\Pi_t(\vek{p})} & = & \frac{- i a_s}{2\pi^2 m \hbar^2} \sum_{l,l'=-\infty}^\infty \psi_l \psi_{l'} \int d^3 p_1 \int d^3 p_1' \nonumber \\ && \times \tilde{\phi}(\vek{p}_1) \tilde{\phi}(\vek{p}_1') \int_0^t dt' e^{i \omega_{l, l'}(\vek{p}-\vek{p}_1,\vek{p}-\vek{p}_1') t'}  \label{eq:Pt}\\ && \times \hat{\psi}^\dagger[(l + l') \hbar k_L \vek{e}_z + \vek{p}_1 + \vek{p}_1' - \vek{p} ] \hat{b}_0^2 \ket{\Psi_0}\nonumber 
\end{eqnarray}
where we define
\begin{equation}
  \omega_{l, l'}(\vek{q},\vek{q}') = \frac{1}{m \hbar} \left( l \hbar k_L \vek{e}_z - \vek{q} \right) \cdot \left( l' \hbar k_L \vek{e}_z - \vek{q}' \right) \,.
\end{equation}
The time integral appearing in Eq.~\eqref{eq:Pt} is straightforwardly evaluated yielding
\begin{equation}
  \int_0^t dt' e^{i \omega_{l, l'}(\vek{q},\vek{q}') t'} = t \, \mathrm{sinc}[\omega_{l, l'}(\vek{q},\vek{q}') t / 2] e^{i \omega_{l, l'}(\vek{q},\vek{q}') t / 2}
\end{equation}
with $\mathrm{sinc}(\xi) \equiv \sin(\xi)/\xi$. It has a similar effect as Dirac's delta distribution for large $t\to \infty$ insofar as it would yield vanishing contributions within Eq.~\eqref{eq:Pt} for values of $\omega_{l, l'}(\vek{q},\vek{q}')$ that are different from zero.
Hence, given the fact that the prefactor $\tilde{\phi}(\vek{p}_1) \tilde{\phi}(\vek{p}_1')$ restrains the effective integration domain of $\vek{p}_1$ and $\vek{p}_1'$ to a very narrow region about the origin, we can infer that for a sufficiently long evolution time $t$ a nonvanishing momentum density can be encountered only in the immediate vicinity of the surface of spheres in momentum space that have their north and south poles at the points $l \hbar k_L \vek{e}_z$ and $l' \hbar k_L \vek{e}_z$ and are therefore characterized by the equation
\begin{equation}
  (\vek{p}-l \hbar k_L \vek{e}_z)\cdot(\vek{p}-l' \hbar k_L \vek{e}_z) = 0
\end{equation}
for any pair of integers $l,l'\in\mathbb{Z}$.

To quantitatively calculate the momentum density according to Eq.~\eqref{eq:np}, we can set
\begin{equation}
  \hat{\psi}[(l + l') \hbar k_L \vek{e}_z + \vek{p}_1 + \vek{p}_1' - \vek{p} ] \ket{\Psi_0} = 0
\end{equation}
by the choice that we made for $\vek{p}$, since we effectively have $| \vek{p}_1 + \vek{p}_1' | \ll \hbar k_L$ according to the above reasoning.
Similarly, we can evaluate
\begin{eqnarray}
   && \left[ \hat{\psi}[(l_2 + l_2') \hbar k_L \vek{e}_z + \vek{p}_2 + \vek{p}_2' - \vek{p} ],\right. \nonumber \\ && \left.\hat{\psi}^\dagger[(l_1 + l_1') \hbar k_L \vek{e}_z + \vek{p}_1 + \vek{p}_1' - \vek{p} ] \right] \nonumber \\ && =\delta_{l_1 + l_1', l_2 + l_2'} \delta(\vek{p}_1 + \vek{p}_1' - \vek{p}_2 - \vek{p}_2')
\end{eqnarray}
when this commutator is multiplied with the prefactor $\tilde{\phi}^*(\vek{p}_2) \tilde{\phi}^*(\vek{p}'_2) \tilde{\phi}(\vek{p}_1) \tilde{\phi}(\vek{p}_1')$. The two spheres on which $\vek{p}$ has to be simultaneously located in order to yield a significant momentum density would therefore have the same center at $\frac12 (l_1 + l_1') \hbar k \vek{e}_z = \frac12 (l_2 + l_2') \hbar k \vek{e}_z$, which implies that they are actually identical and we have either $l_1 = l_2$ and $l_1' = l_2'$ or $l_1 = l_2'$ and $l_1' = l_2$.

Using $\bra{\Psi_0}\hat{b}_0^\dagger\hat{b}_0^\dagger\hat{b}_0\hat{b}_0\ket{\Psi_0} = N(N-1) \simeq N^2$ for $N\gg 1$ and
\begin{equation}
  \delta(\vek{p}_1 + \vek{p}_1' - \vek{p}_2 - \vek{p}_2') = \int \frac{d^3 r}{(2\pi \hbar)^3} e^{i(\vek{p}_1 + \vek{p}_1' - \vek{p}_2 - \vek{p}_2')\cdot \vek{r} / \hbar}
\end{equation}
we can, consequently, write the momentum density \eqref{eq:np} as
\begin{eqnarray}
  n(\vek{p},t) & \simeq & \sum_{l,l'=-\infty}^\infty |\psi_l|^2 |\psi_{l'}|^2 \nonumber \\ && \times n\left(\vek{p} - \frac{l + l'}{2} \hbar k \vek{e}_z, \frac{l - l'}{2} \hbar k \vek{e}_z, t \right)
\end{eqnarray}
with
\begin{equation}
  n(\vek{p},\vek{p}_0,t) = \int d^3r | \chi(\vek{r},\vek{p},\vek{p}_0,t) |^2
\end{equation}
where
\begin{eqnarray}
  \chi(\vek{r},\vek{p},\vek{p}_0,t) & = & \frac{N a_s}{4\sqrt{\pi\hbar}^7 m} \int d^3 p_1 \int d^3 p'_1   \nonumber \\ &&\times  \int_0^t dt'e^{i t'(\vek{p}_1 + \vek{p}_0 - \vek{p})\cdot(\vek{p}'_1 - \vek{p}_0 - \vek{p} ) / (m \hbar)} \nonumber \\ &&\times \tilde{\phi}(\vek{p}_1) \tilde{\phi}(\vek{p}'_1) e^{i (\vek{p}_1 + \vek{p}'_1)\cdot \vek{r} / \hbar} \label{eq:chi}
\end{eqnarray}
represents some sort of phase-space wavefunction that describes the collision process between two momentum components $l$ and $l'$ of the condensate wavefunction, with $\vek{p}_0 = \frac{1}{2}( l - l') \hbar k_L \vek{e}_z$ and with the origin in momentum space being set at $\frac{1}{2}( l + l') \hbar k_L \vek{e}_z$.
In the context of scattering halos, we are specifically interested in the case $l \neq l'$, which implies a finite relative momentum $2p_0 \sim \hbar k_L$ of the two wavefunction components.
We can therefore neglect the $\vek{p}_1\cdot\vek{p}'_1$ term arising in the exponent within Eq.~\eqref{eq:chi} as it will become negligibly small with respect to the other terms in this exponent, owing to the presence of the prefactor $\tilde{\phi}(\vek{p}_1) \tilde{\phi}(\vek{p}'_1)$.
This consequently yields
\begin{eqnarray}
  \chi(\vek{r},\vek{p},\vek{p}_0,t) & \simeq & \frac{2 N a_s}{\sqrt{\pi\hbar} m} \int_0^{t} d t' e^{i t'(\vek{p} - \vek{p}_0)\cdot(\vek{p} + \vek{p}_0)/(m\hbar)} \nonumber \\ && \times \phi[\vek{r} - t'(\vek{p} + \vek{p}_0)/m] \nonumber \\ && \times\phi[\vek{r} - t'(\vek{p} - \vek{p}_0)/m] \,.
\end{eqnarray}

Making use of the fact that the Thomas-Fermi profile \eqref{eq:TF} characterizing the condensate wavefunction is of finite extent, we can safely take the limit $t\to\infty$ in the above expression and obtain
\begin{eqnarray}
  n(\vek{p},\vek{p}_0) & \simeq & \frac{4 N^2 a_s^2}{\pi \hbar p_0^2 R} \int_0^\infty d\tau \int_0^\infty d\tau' \label{eq:npp0} \\ &&\times \exp\left[ \frac{i R}{\hbar p_0}( p^2 - p_0^2)(\tau - \tau')\right] \int d^3\rho \nonumber \\ &&\times \varphi[\veg{\rho} - \tau( \vek{p} / p_0 + \vek{e}_z)]\varphi[\veg{\rho} - \tau( \vek{p} / p_0 - \vek{e}_z)] \nonumber \\ && \times \varphi[\veg{\rho} - \tau'( \vek{p} / p_0 + \vek{e}_z)]\varphi[\veg{\rho} - \tau'( \vek{p} / p_0 - \vek{e}_z)] \nonumber
\end{eqnarray}
for the momentum density of scattered atoms in the long-time limit, where we use $\vek{p}_0 = \pm p_0 \vek{e}_z$ with $p_0 = |\vek{p}_0|$.
This expression can be further simplified by using the fact that for $p \neq p_0$ the term $\exp[i R (p^2 - p_0^2)(\tau-\tau') / (\hbar p_0)]$ has, in the limit $R p_0 / \hbar \to \infty$, a similar effect as a delta function in $\tau - \tau'$ within Eq.~\eqref{eq:npp0}.
We are therefore entitled to rewrite this expression as
\begin{eqnarray}
  n(\vek{p},\vek{p}_0) & \simeq & \frac{8 N^2 a_s^2}{\pi p_0 R^2} \int_0^\infty d\tau \frac{\sin[\tau R( p^2 - p_0^2) /(\hbar p_0)]}{p^2 - p_0^2} \nonumber \\ && \times \int d^3\rho\varphi^2(\veg{\rho} - \tau \vek{e}_z) \varphi^2(\veg{\rho} + \tau \vek{e}_z) \,, \label{eq:halo}
\end{eqnarray}
which for $R p_0 / \hbar \to \infty$ yields a tight and isotropic concentration of the momentum density around the sphere with the radius $p = p_0$.
As is clearly seen from Eq.~\eqref{eq:halo}, the width of these scattering halos in momentum space is of the order of $\Delta p \sim \hbar / R$, which is in agreement with the experimental findings.

\begin{figure}[t]
  \includegraphics[width=\linewidth]{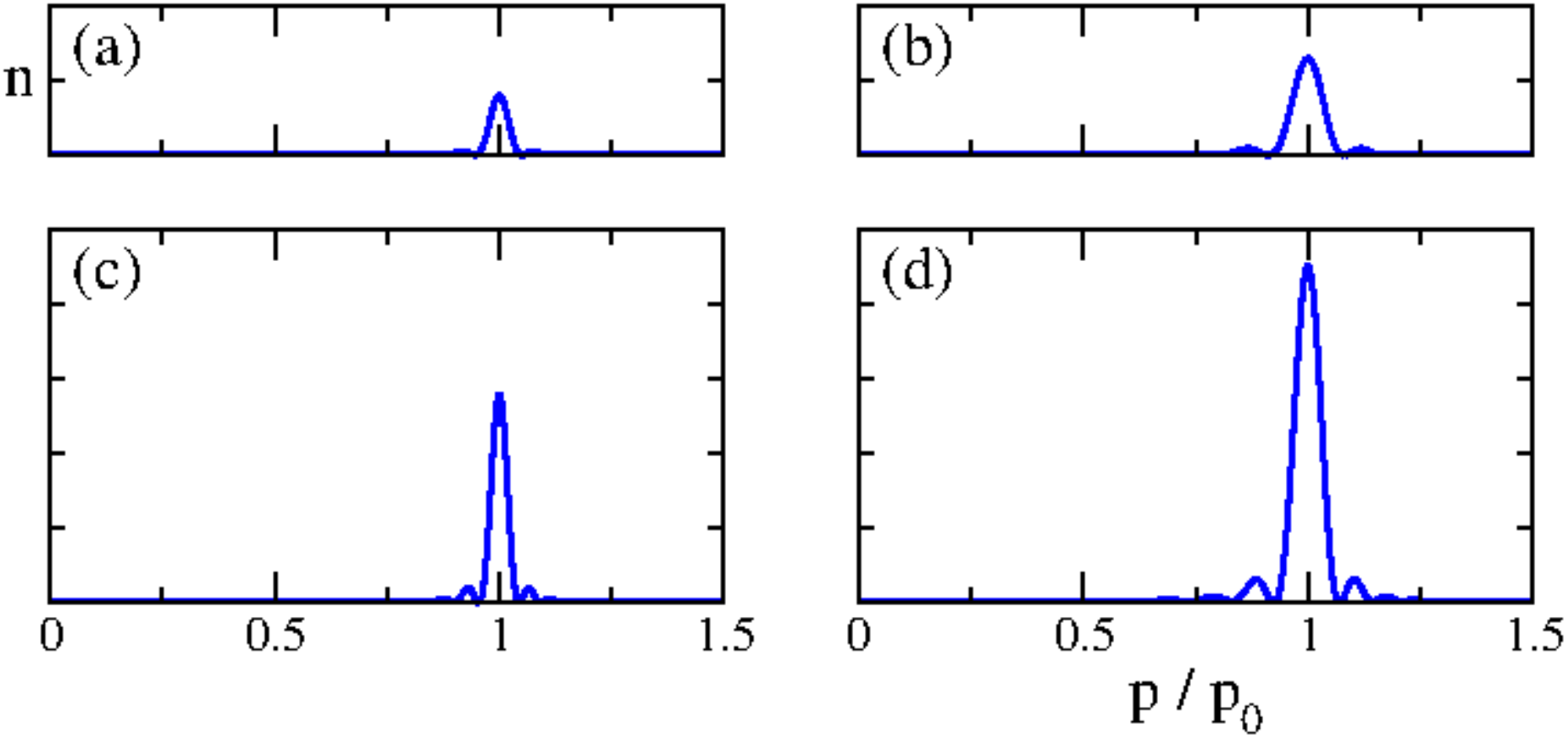}
  \caption{\label{fig:tof} Momentum density profiles (plotted in arbitrary but for all panels fixed units) as a function of $p = |\vek{p}|$ for $p_0 = \hbar k_L / 2$ (corresponding to $|l-l'|=1$) with $k_L = 2\pi/d$ and $d = 532\,\mathrm{nm}$. The profiles are obtained from a numerical evaluation of Eq.~\eqref{eq:halo} for the case of $^{87}$Rb, where we assume the presence of a spherical trap ($\omega_\perp = \omega_{||}= \bar{\omega}$) with the confinement frequencies (a,c) $\bar{\omega} = 2\pi\times 30\,\mathrm{Hz}$ and (b,d) $\bar{\omega} = 2\pi\times 100\,\mathrm{Hz}$, and where we consider a condensate population containing (a,b) $N = 10^5$ and (c,d) $N = 2\times 10^5$ atoms. The resulting Thomas-Fermi radii are evaluated as (a) $R \simeq 10\,\mu\mathrm{m}$, (b) $R \simeq 6.4\,\mu\mathrm{m}$, (c) $R \simeq 12\,\mu\mathrm{m}$, (d) $R \simeq 7.4\,\mu\mathrm{m}$. The full width at half maximum (FWHM) of the momentum density peaks equals roughly $\pi \hbar / R$. }
\end{figure}

This is confirmed in Figure \ref{fig:tof} which shows a numerical evaluation of Eq.~\eqref{eq:halo} for the experimental parameters under consideration and for various choices of the condensate population and the overall confinement frequency (assuming a spherical trap). The atomic density in momentum space is concentrated about a sphere of radius $p_0 = |l-l'| \hbar k_L / 2 = \hbar k_L / 2$ in this example ($|l-l'|=1$). The width (FWHM) of this scattering halo is found to roughly equal $\pi \hbar / R$ in all of the four studies cases. Note that these density profiles cannot be directly compared with the profile shown in Fig.~\ref{fig:quantification}\textbf{b} since the latter is the result of the integration of the scattering sphere along the imaging axis, followed by angular averaging.

The total number of atoms that are scattered as a consequence of the collision between the $l$ and $l'$ components of the condensate wavefunction (with $l \neq l'$) is then straightforwardly calculated as $N_{\rm coll. at.} = 2|\psi_l|^2 |\psi_{l'}|^2 \mathcal{N}$ with
\begin{eqnarray}
  \mathcal{N} & = & \int d^3p n(\vek{p},\vek{p}_0) \nonumber \\ & \simeq & \frac{16 N^2 a_s^2}{R^2} \int_0^\infty d\tau \int d^3\rho\varphi^2(\veg{\rho} - \tau \vek{e}_z) \varphi^2(\veg{\rho} + \tau \vek{e}_z) \nonumber \\ & = & \left[ \frac{5 N a_s}{2 R} \left(\frac{\omega_\perp}{\omega_{||}}\right)^{1/3}\right]^2 \, \label{eq:N} \\ & = & \frac{\sigma}{8\pi R^2}\left[ \frac{5 N}{2} \left(\frac{\omega_\perp}{\omega_{||}}\right)^{1/3}\right]^2 \,. \label{eq:crosssection} 
\end{eqnarray}
In this last expression we have isolated the scattering cross section $\sigma$, which is here independent of the colliding orders $l$ and $l'$. Eq.~\ref{eq:N} is quantitatively rather similar to the analogous Eq.~(2) for the number of collisions within Ref.~\cite{Tenart2020}, which was obtained there in the framework of a three-dimensional optical lattice.
The scaling with $(\omega_\perp/\omega_{||})^{2/3}$ translates the fact that two cigar-shaped condensates moving across each other along their symmetry axis give rise to more atom-atom collisions than two pancake-shaped condensates. For a mean angular frequency $\bar \omega = 2\pi \times 50 $ Hz, a number of atoms equal to $2\times 10^5$, and an ideal, equal-weights collision, $|\psi_l|^2=|\psi_{l'}|^2=0.5$, we find a fraction $N_{\rm coll. at.}/N\simeq 20$~\% of atoms that collide. This is somewhat smaller than what is observed experimentally, which may indicate that a non-perturbative approach is required for a quantitative agreement.

\section{Wavefunction calculation of the distribution of orders of diffraction}
\label{appendixC}

The eigenstates of the lattice hamiltonian with the potential Eq.~\ref{eq:potential} are Bloch functions of the form :

\begin{align}
\ket{\Phi_{n,q}}=\sum_{l\in \mathds{Z}} C_l^{(n,q)} \ket{\chi_{q+lk_L}}, \nonumber
\end{align}
where the vectors $\ket{\chi_k}$ are eigenstates of the momentum operator with eigenvalue $\hbar k$ (and $\chi_k(x)=\frac{e^{ikx}}{\sqrt{2\pi}}$), the quasi-momentum $q\in [-\pi/d,\pi/d[$, and the coefficients $C_l^{(n,q)}$ are the solution of the \emph{central equation} (for a given quasi-momentum $q$, and $\varphi=0$) \cite{morsch2006}:
\begin{align}
\frac{1}{4}\left(2l+\frac{q}{k_L}\right)^2C_l^{(n,q)}-\frac{s}{4}\left(C_{l+1}^{(n,q)}+C_{l-1}^{(n,q)}\right)=\frac{E_{n,q}}{E_L} C_l^{(n,q)}.
\end{align}
This equation can easily be solved in matrix form to obtain the eigenenergies of the system and the decomposition of the eigenstates on the plane wave basis.

The BEC is initially loaded in the lowest energy eigenstate of the lattice, corresponding to a quasi-momentum $q=0$ : $\ket{\Psi(t=0})=\ket{\Phi_{0,0}}$. After the sudden phase shift, the wavefunction in the reference frame of the lattice takes the form :
\begin{align}
\ket{\Psi(t=0^+)}=\sum_{l\in \mathds{Z}} C_l^{(0,0)} e^{-il\varphi_0} \ket{\chi_{lk_L}}. \nonumber
\end{align}

Since the lattice preserves its periodicity in the shift, the quasi-momentum is conserved. The wavefunction after the shift can therefore be decomposed onto the eigenstates of the shifted lattice with quasi-momentum $q=0$
to obtain the evolution during the holding time $t_\mathrm{hold}$ :
\begin{align}
\ket{\Psi(t_\mathrm{hold})}=\sum_{n\in \mathds{N}} \alpha_n e^{-i\frac{E_{n,0}t_\mathrm{hold}}{\hbar}}\ket{\Phi_{n,0}}, \nonumber
\end{align}
with $\alpha_n=\bracket{\Phi_{n,0}}{\Psi(t=0^+)}=\sum_{l\in \mathds{Z}} C_l^{*(n,0)}C_l^{(0,0)} e^{-il\varphi_0}$.

Finally the population in the diffraction order of momentum $j\times h/d$ is obtained by calculating the projection of the state $\ket{\Psi(t_\mathrm{hold})}$ on the corresponding eigenstate of the momentum operator $\ket{\chi_{jk_L}}$:
\begin{align}
\pi_j=\left|\bracket{\chi_{jk_L}}{\Psi(t_\mathrm{hold})}\right|^2=\left|\sum_{n\in \mathds{N}} \alpha_n C_j^{(n,0)} e^{-i\frac{E_{n,0}t_\mathrm{hold}}{\hbar}}\right|^2. \nonumber
\end{align}

\section{Impact of in-lattice dynamics on the collisional halos}
\label{appendixB}

\begin{figure*}[ht!]
	\includegraphics{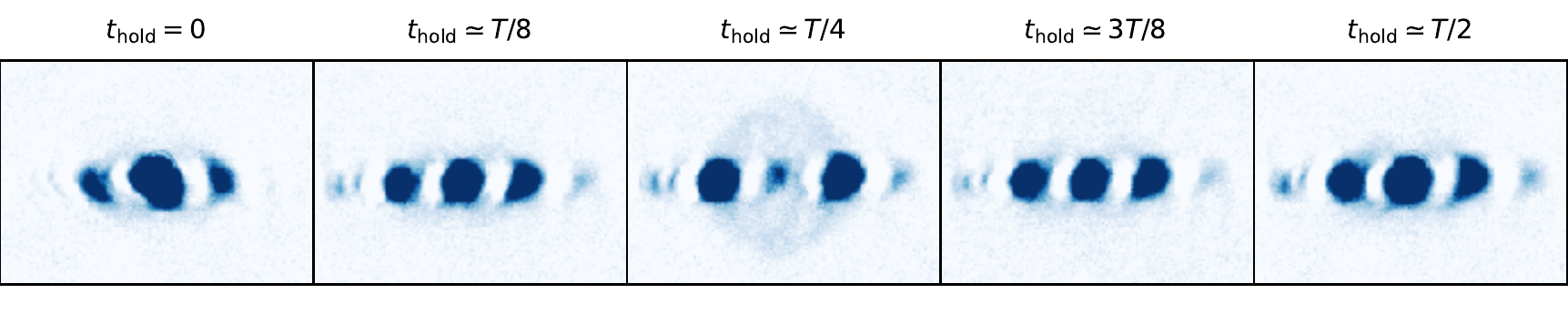}
	\caption{Single-shot images of diffraction orders for varying holding times $t_\mathrm{hold}$ expressed as a fraction of the dipolar motion period $T$ (here $T\simeq\SI{140}{\micro\second}$), for a shift of $\varphi_0=\ang{180}$, and a lattice depth of $s=1.92\pm0.07$. The color scale is set to enhance the collision halos (clipping high values on some diffraction orders). The collisional halo between orders $\pm1\times \hbar k_L$ is only observed when these orders are populated at the time of release from the lattice, for $t_\mathrm{hold}\simeq T/4$.}
	\label{fig:evolution}
\end{figure*}

In this appendix we show that the collisional halos observed in this article are determined only by the momentum distribution at the end of the out-of-equilibrium dynamics in the lattice, of duration $t_\mathrm{hold}$. This is illustrated in figure \ref{fig:evolution}, where we have represented the collision halos recorded for a phase shift $\varphi_0=180^{\textrm{\tiny{o}}}$ and a lattice depth of $s=1.92\pm0.07$, for various durations $t_\mathrm{hold}$. These are expressed as a fraction of the period $T$ of the dipolar motion in the lattice (here $T\simeq\SI{140}{\micro\second}$). 

In the first couple images, no collision halo between orders $\pm1\times \hbar k_L$ can be seen, and it only appears for $t_\mathrm{hold}=T/4$, when these orders are significantly populated. As the dynamics is extended further, and the final population in orders $\pm1\times \hbar k_L$ reduces again, the corresponding halo is not observed anymore. If collisions were occurring inside the lattice, the collision halo would remain visible for all holding times after its first appearance.
This demonstrates that the halos are determined by the momentum distribution upon release from the lattice, and not affected by the prior dynamics.

\end{appendix}

\bibliographystyle{apsrev4-1}
\bibliography{collisionsAll}
\end{document}